\documentclass{cernrep}
\usepackage{texnames}
\usepackage[T1]{fontenc}

\usepackage{varwidth}
\usepackage{xcolor}

\begin{document}
\title{Microwave Coupling to
ECR and Alternative Heating Methods}

\author{L. Celona}

\institute{Istituto Nazionale di Fisica Nucleare, Laboratori Nazionali del Sud, Catania, Italy}

\maketitle 

\begin{abstract}
The Electron Cyclotron Resonance Ion Source (ECRIS) is nowadays
the most effective device that can feed particle accelerators in a continuous
and reliable way, providing high-current beams of low- and medium-charge-state ions
and relatively intense currents for highly charged ions. The ECRIS is an
important tool for research with ion beams (in surface, atomic, and nuclear science) while,
 on the other hand, it implies plasma under extreme conditions and thus constitutes an
 object of scientific interest in itself. The fundamental aspect of the coupling between the electromagnetic wave and the plasma is hereinafter treated together with some variations to the classical ECR heating mechanism, with particular attention being paid to the frequency
tuning effect and two-frequency heating.
Considerations of electron and ion dynamics will be presented together with some recent observations connecting the beam shape with the frequency of the electromagnetic wave feeding the cavity.
The future challenges of higher-charge states, high-charge breeding efficiency, and high absolute ionization efficiency also call for the exploration of new heating schemes and synergy between experiments and modelling. Some results concerning the investigation of innovative mechanisms of plasma ignition based on upper hybrid resonance will be described.


\end{abstract}

\section{Introduction}
The Electron Cyclotron Resonance Ion Source (ECRIS) is used to
deliver beams of singly or multiply charged ions for a
wide range of applications in many laboratories. In particular,
such devices are well suited for the production of Highly
Charged Ions (HCIs) -- a key consideration for the new acceleration
facilities (e.g. FAIR, RIA, HRIBF, etc.), which
will require milliampere levels of HCIs. In order to get such high
currents, the recent progress in the performance of the ECRIS has
been mainly linked to improvement of the plasma magnetic
confinement within the source chamber and to an increase of the
frequency for the feeding microwaves which, under
proper conditions, develop to higher plasma densities \cite{gam}.
The trend followed until now has been to increase the frequency and the magnetic
field, leading to rising costs for the technology and safety problems for the magnet's
cryostat because of the growth of hot electrons: the frequency and
magnetic field scaling are close to saturation.
Apart from this main route, various techniques are also
employed to enhance the production of HCIs, such as the use of
secondary emission materials, the wall coatings, the installation
of a bias disc, or gas mixing, to mention the most important ones \cite{dren}.

Moreover, in the past decade, some experiments involving
new approaches to feeding have allowed an increase in the performance
of the conventional ECR ion sources with a
minimum-$B$ magnetic field structure by feeding them with
electromagnetic waves that have a large spectral content, or that are obtained
by the superimposition of a discrete set of microwaves
at different frequencies \cite{kaw,xie,vonf,alton}.

Even if these experiments have provided interesting results,
they have not given us an explanation or a methodology to convey a
better understanding of the coupling mechanism between a feeding
waveguide and a cavity filled with plasma, and the energy
transfer between the electromagnetic field in the source
plasma chamber and the plasma confined therein.
An increase of knowledge in terms of microwave coupling to
ECRIS plasma, and therefore optimization of the ECR
power transfer processes, may allow us to design ion sources
with higher performance.
Further improvements of ECRIS output currents and the average
charge state require a deep understanding of electron and ion dynamics in the plasma, and of the impact of the electromagnetic structure on the electron density distribution:
theoretical investigations, in order to find a reasonable and adequate model for these mechanisms and to predict their possible improvement, may lead to the design of a future innovative ECR ion source.

Over recent years, dedicated experiments at
INFN-LNS have aimed to investigate these topics and, at the
same time, some theoretical studies have been undertaken to
describe them in detail \cite{celona}.

The microwave coupling between the electromagnetic wave and the plasma
determines the efficient transfer of the energy from the microwaves to the
plasma electrons inside the ECRIS. The plasma chamber, on the other hand,
can be considered as a resonant cavity for the electromagnetic waves and, when the magnetized
plasma is created, the electrical permittivity of the medium filling the chamber
is no longer homogeneous and no longer electrically isotropous. A detailed investigation has been carried out to demonstrate that the performance
of ECR ion sources depends on the electromagnetic field excited inside
the plasma chamber and on the coupling mechanism used to provide the
microwaves to the plasma.
In the following, the various approaches used over the years are presented, together with the possible alternative heating schemes.

%
%

\section{The ECR Standard Model}
Up to now, the so-called `ECRIS Standard Model' has, for the greater part of
a decade, been the road map followed by ECRIS designers.
The main rules were confirmed by experiments performed
at MSU-NSCL in 1993--4 and 1995, and can be
summarized as follows \cite{msu}:
\begin{itemize}
\item the radial magnetic field value at the plasma chamber wall must be
$B_{\mathrm{rad}}$ $\geq$ $2B_{\mathrm{ECR}}$;
\item the axial magnetic field value at injection must be
$B_{\mathrm{inj}}$ $\simeq$ $3B_{\mathrm{ECR}}$ or more;
\item the axial magnetic field value at extraction must be about
$B_{\mathrm{ext}}$ $\simeq$ $B_{\mathrm{rad}}$;
\item the minimum value of the axial magnetic field must be in the range
$0.30<B_{\mathrm{min}}/B_{\mathrm{rad}}<0.45$; and
\item the optimum power must increase with the volume of the plasma and with
the square of the frequency.
\end{itemize}
Up to now, almost every operating ECRIS has complied with the Standard Model: the extracted current
increases as the microwave frequency increases, but only an increase in the
mirror ratio can exploit the optimal performance, making the increase of the
electron density with frequency effective.
Also, according to the Standard Model, the development of the ECRIS is strictly
linked to improvements in superconducting magnets and in the technology of microwave generation.
Various authors have studied the RF coupling to the plasma in terms of the
maximum power rate per unit volume and its relationship with the beam
intensity produced by different ECR ion sources \cite{source}, but this description is
not satisfactory. The cavity design, and the microwave injection geometry,
are of primary importance for high-RF-energy transmission to the plasma
chamber. Note that the problem of wave energy transmission into the
plasma must be divided into two parts: the first is related to the microwave generator--waveguide--plasma chamber coupling, while the second concerns
the wave--plasma interaction. Both of these two aspects play notable roles for the
future improvement of ECRIS performance.
Many experiments have been undertaken outside the framework traced by the Standard
Model, to verify the possibility of improving the plasma heating. Principally, they follow
three different road maps. A first series of experiments has been devoted to the study
of the variation of ECRIS performance with slight variations of the microwave
frequency. The second route regards the possibility of operating with more
than one frequency for plasma heating (usually two) or of using broad microwave spectra. The
third method is actually a mixture of the previous ones: two or more frequencies
are used for plasma heating, but at least one is provided by a broadband
microwave generator (such as a Travelling-Wave Tube, or TWT), which allows the effects of
frequency tuning and of multi-frequency heating to be combined.

\subsection{The ECRIS plasma chamber as a resonant cavity}
In order to characterize the electromagnetic field that is present inside
the plasma chamber of an ECR ion source and where the particle motion
occurs, it is fundamental to make the following assumption: the plasma
chamber is a resonating cavity for the electromagnetic wave feeding the
plasma. The coupling between the electromagnetic wave and the plasma-filled
chamber and the electromagnetic field patterns that can be excited
inside the resonating cavity are of primary importance for the characterization
and for the properties of the plasma. Furthermore, the characterization
of the plasma chamber in terms of the excitable electromagnetic field allows us to
make some assumptions about the charged particle motion and energy, as
described in the following sections.
\subsection{Resonant modes inside a cylindrical plasma chamber}
A plasma chamber can be represented, in a first-order approximation, by a
cylinder of radius \emph{a} and length \emph{l}, filled by a medium with a certain electric
permittivity $\epsilon$ and a certain magnetic permeability $\mu$. Then, a discrete
number of electromagnetic field patterns can exist inside the plasma chamber:
the so-called resonant modes. They can be defined by the following equations,
defined using a system of cylindrical coordinates ($\rho$,$\phi$,$z$):
\begin{equation}
\label{eqn:risonanza_tm}
\begin{cases}
E_\rho = -\displaystyle\frac{x_{n\nu}}{a}\displaystyle\frac{r}{l}\displaystyle\frac{\pi}{h^2}C_nJ'_n\left(\frac{x_{n\nu}\rho}{a}\right)\cos{n\phi}\sin{\left({\displaystyle\frac{r\pi z}{l}}\right)},\\
E_\phi = \displaystyle\frac{n}{\rho}\displaystyle\frac{r}{l}\displaystyle\frac{\pi}{h^2}C_nJ_n\left(\displaystyle\frac{x_{n\nu}\rho}{a}\right)\sin{n\phi}\sin\left({\displaystyle\frac{r\pi z}{l}}\right),\\
E_z = C_nJ_n\left(\displaystyle\frac{x_{n\nu}\rho}{a}\right)\cos{n\phi}\cos\left({\displaystyle\frac{r\pi z}{l}}\right),\\
H_\rho = -i\displaystyle\frac{\epsilon\omega}{h^2}\displaystyle\frac{n}{\rho}C_nJ_n\left(\displaystyle\frac{x_{n\nu}\rho}{a}\right)\sin{n\phi}\cos\left({\displaystyle\frac{r\pi z}{l}}\right),\\
H_\phi = -i\displaystyle\frac{\epsilon\omega}{h^2}\displaystyle\frac{x_{n\nu}}{a}C_nJ'_n\left(\displaystyle\frac{x_{n\nu}\rho}{a}\right)\cos{n\phi}\cos\left({\displaystyle\frac{r\pi z}{l}}\right),\\
H_z = 0.
\end{cases}
\end{equation}

These are the equations describing the Transverse Magnetic (TM) modes, with only
the magnetic field components on the transverse plane. The Transverse Electric (TE)
modes, with only the electric field components on the transverse plane, can be calculated
using the following equations:
\begin{equation}
\label{eqn:risonanza_te}
\begin{cases}
E_\rho = \displaystyle\frac{\mu\omega}{h^2}\displaystyle\frac{n}{p}C_nJ_n\left(\displaystyle\frac{x'_{n\nu}\rho}{a}\right)\sin{n\phi}\sin\left({\displaystyle\frac{r\pi z}{l}}\right),\\
E_\phi = \displaystyle\frac{\mu\omega}{h^2}\displaystyle\frac{x_{n\nu}}{a}C_nJ'_n\left(\displaystyle\frac{x'_{n\nu}\rho}{a}\right)\cos{n\phi}\sin\left({\displaystyle\frac{r\pi z}{l}}\right),\\
E_z = 0\\
H_\rho = -i\displaystyle\frac{x'_{n\nu}}{a}\displaystyle\frac{r}{l}\displaystyle\frac{\pi}{h^2}C_nJ'_n\left(\displaystyle\frac{x_{n\nu}\rho}{a}\right)\cos{n\phi}\cos\left({\displaystyle\frac{r\pi z}{l}}\right),\\
H_\phi = i \displaystyle\frac{n}{\rho}\displaystyle\frac{r}{l}\displaystyle\frac{\pi}{h^2}C_nJ_n\left(\displaystyle\frac{x'_{n\nu}\rho}{a}\right)\sin{n\phi}\cos\left({\displaystyle\frac{r\pi z}{l}}\right)\\
H_z = -iC_nJ_n\left(\displaystyle\frac{x'_{n\nu}\rho}{a}\right)\cos{n\phi}\sin\left({\displaystyle\frac{r\pi z}{l}}\right).\\
\end{cases}
\end{equation}
Here, the time dependence $\mathrm{e}^{i\omega t}$ is omitted; $J_n$ and $J'_n$ are, respectively, the Bessel functions and the derivatives of the Bessel functions of order $n$;  and $x_{n\nu}$ and $x'_{n\nu}$ are their related $\nu$ roots. A discrete set of frequencies $\omega$/2$\pi$ can exist inside the resonance cavity and they are defined as follows:

\begin{equation}
\label{eqn:omega}
\omega_{n\nu r}=c\sqrt{\frac{r^2\pi^2}{l^2}+h^2},
\end{equation}

where the value $h$ is as follows:
\begin{alignat}{4}
\label{eqn:te}
h=\frac{x'_{n\nu}}{a} \hspace{1cm} (\mathrm{TE \hspace{0.2cm} modes}),\\
\,\,=\frac{x_{n\nu}}{a} \hspace{1cm} (\mathrm{TM \hspace{0.2cm} modes}).
\end{alignat}
Then, the resonant
modes TE$_{n,\nu,\rho}$ and TM$_{n,\nu,\rho}$ that exist inside the plasma chamber are identified
by the mode parameters and their strength is
related to the amplitude associated with the field strength of the wave.
By supposing that the chamber is filled with a homogeneous
and unmagnetized medium, the corresponding electrical
permittivity is given by the relation
\begin{equation}
\label{eqn:epsi}
\epsilon_{r}=1-\left(\frac{\omega_p}{\omega}\right)^2,
\end{equation}
where $\omega$ is the pulsation generating the plasma and $\omega_p$ is the plasma pulsation:
\begin{equation}
\label{eqn:omega_p}
\omega_{p}=\sqrt{\frac{n_\mathrm{e}e^2}{m_\mathrm{e}\epsilon_o}},
\end{equation}
in which $m_\mathrm{e}$ and $e$ are, respectively, the electron mass and the
electron charge, $\epsilon_o$ is the electrical permittivity in vacuum,
and $n_\mathrm{e}$ is the electron density. Therefore, under these hypotheses,
the plasma build-up can be seen as a change in
the electrical permittivity and therefore as a change in the
resonant frequencies, because of the change of $\epsilon$ in Eq. (\ref{eqn:omega}).
This model is certainly oversimplified if we consider the plasma of an ECR ion source, but has been applied successfully, giving information on the plasma parameters generated in a small reactor at INFN-LNS.

\section{Two- and multiple-frequency heating}
Since 1994, the so-called Two-Frequency Heating (TFH) has been used \cite{xie,vonf} to improve the HCI
production by feeding the plasma with two electromagnetic waves at different frequencies instead of one. In some cases, even three or more close
frequencies have been used.

The performance of the LBL AECR source has been improved by simultaneously heating the plasma
with microwaves of 10 and 14 GHz. The plasma stability was improved and
the ion charge-state distribution was shifted to a higher-charge state. The production of
high-charge-state ions was increased by a factor of between 2 and 5, or higher, for the very heavy ions
such as bismuth and uranium, as
compared to the case of single-frequency (14 GHz) heating \cite{xie}.

In an ECR source, electron cyclotron resonance heating couples microwave power into the
plasma electrons. This occurs when the microwave frequency $\omega_{rf}$ matches the cyclotron frequency,
$\omega_c=B/m_\mathrm{e}$, of the electrons. In high-charge-state ECR sources with one frequency, the geometry
of the minimum-$B$ field results in a closed, approximately ellipsoidal ECR surface. The electrons
are heated in a thin resonance zone at the surface, as they spiral back and forth between the magnetic
mirrors. When two frequencies are used, it is possible to produce two concentric surfaces, the
physical separation of which depends on the frequency difference and the strength and gradient of the
magnetic field, leading to a higher density of the energetic
electrons.
With TFH, the ECR plasma is more
quiescent than with single-frequency heating. Both the short-term and the long-term plasma stability
are improved, and more microwave power could be launched into the plasma.
TFH has been demonstrated to be a powerful method: in the case in Ref. \cite{xie} for
$^{238}$U, it increased the production of higher-charge states (from 35+ to 39+)
by a factor ranging from 2 to 4 and shifted the peak charge state from 33+ to
36+.
Two-frequency heating using 10 and 14 GHz in the AECR provided significantly better
performance and indicated that still higher performance with multiple-frequency heating may be
possible. While multiple-frequency heating would increase the complexity and cost of an ECR
source, it could provide significant gains in performance.
In \cite{vonf}, the performance of the Argonne ECR ion source was improved through the use of TFH, with the primary frequency of 14 GHz from a klystron and the second frequency from a Travelling-Wave Tube Amplifier (TWTA) with a tunable range of 11.0--13.0 GHz. Source output as well as stability were improved, with a shift to higher-charge states observed. The use of a second frequency increased the intensity of the medium charge states by 50--100$\%$ and the
higher-charge states by a factor of between 2 and 5.
In a modern second-generation ECRIS, TFH is usually implemented by using frequencies between 14 and 18 GHz, while the best-performing ECRIS use a combination of 24 or 28 GHz with 18 GHz.

\subsection{Multiple-frequency plasma heating}
\begin{figure}[htpb]
\begin{center}
\includegraphics[width=0.5\textwidth]{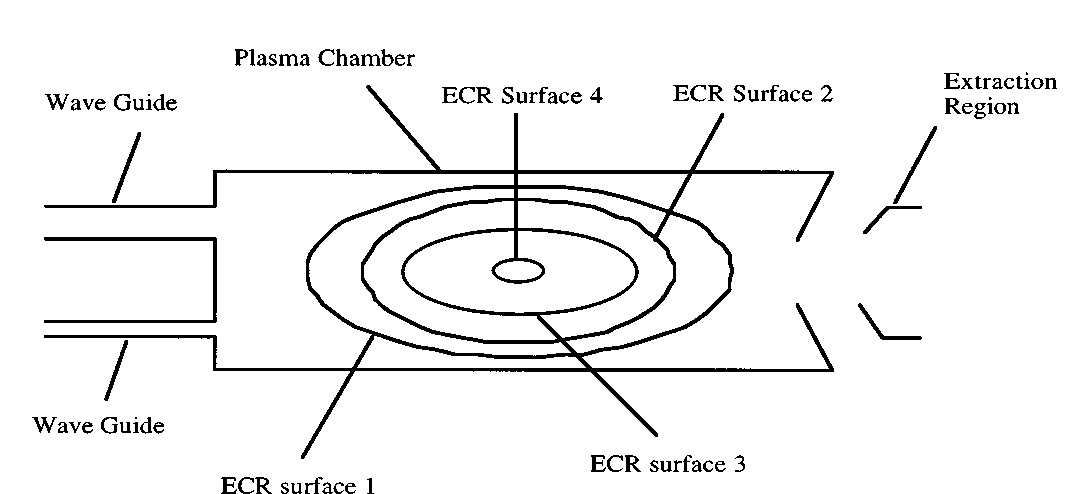}
\end{center}
\caption{A schematic view of four nested ECR surfaces in a high-charge-state
ECR ion source for four well-separated frequency waves \cite{xie2}.\label{mult}}
\end{figure}

Microwaves of various frequencies can be simultaneously
launched into and absorbed by a high-charge-state
ECR plasma. The minimum-$B$ magnetic field configuration
in an ECRIS can provide many closed and nested ECR heating
surfaces, as graphically shown in Fig. \ref{mult}.
If two or more significantly different frequencies are used, two or more
well-separated and nested ECR surfaces
will exist in the ECR plasma. With the multiple ECR surfaces,
electrons can be heated four times or more for one
pass from one mirror end to the other, whereas they are only
heated twice in the case of single-frequency heating.
Multiple-frequency heating can couple more microwave
power with better efficiency into the plasma and it leads to a
higher density of hot electrons, which is essential for the production
of highly charged ions.
To demonstrate that the frequency-domain technique can be used to enhance the performance of a traditional minimum-$B$ ECR ion source, comparative studies have been undertaken to assess the relative performance of the Oak Ridge National Laboratory CAPRICE ECR ion source \cite{maie}, in terms of its multiply charged ion-beam generation capability, when excited with high-power, single-frequency, or multiple-discrete-frequency microwave radiation, derived from standard klystron and/or TWT technologies. These studies demonstrate that the charge-state populations for Ar$^{q+}$ and Xe$^{q+} $move towards higher values when excited with two- and three-discrete-frequency microwave power compared to those observed when single-frequency microwave power is used. For example, the most probable charge state for Xe is increased by one charge-state unit, while the beam intensities for charge states higher than the most probable one are increased by factors of $\sim3$ compared to those observed for single-frequency plasma excitation. In Ref. \cite{com}, particle-in-cell codes have been used to simulate the magnetic field distributions, to demonstrate the advantages of using multiple, discrete frequencies over single frequencies to power conventional ECR ion sources.

One, two and three-frequency heating experiments
have been conducted using Xe feed gas.
With the addition of the second and third frequencies, the
most probable Xe$^{q+}$ charge state moves towards higher values
by one unit; this clearly
 illustrates that the performance of conventional-geometry
ECR ion sources can be significantly improved by the use of
multiple-discrete-frequency plasma heating.

\section{Electron heating with broadband microwave radiation and the volume effect}
The ionization process in Electron Cyclotron Resonance
(ECR) ion sources is based on the sequential removal of
electrons in collisions between ions or neutral atoms with
electrons that have been heated stochastically to high
energies by the adsorption of microwave power under ECR
conditions. The ECR zones in a conventional minimum-$B$
geometry ECRIS, powered by narrow-bandwidth
microwave radiation, are thin volumes that are substantially smaller
in relation to their total plasma volumes. Consequently,
the probability of an acceleration resulting in a substantial
energy gain for the electrons that arrive in the ECR zone in
phase with the electric field vector of the electromagnetic
wave is lower than is possible in extended-volume ECR
zones. Therefore, it is reasonable to believe that the performance
of conventional minimum-$B$ ECR ion sources
can be improved by increasing their respective resonance
volumes. The volume effect has been demonstrated by
tailoring the central magnetic field so that it forms a large
resonant volume  and by increasing the number of
discrete operational frequencies of the ion source.
Broadband sources of RF power provide a simple and
potentially more effective alternative for increasing the
physical sizes of the resonance zones in conventional $B_{\mathrm{minimum}}$
ECR ion sources.
\subsection{Experimental procedures}
\begin{figure}[htpb]
\begin{center}
\includegraphics[width=0.48\textwidth]{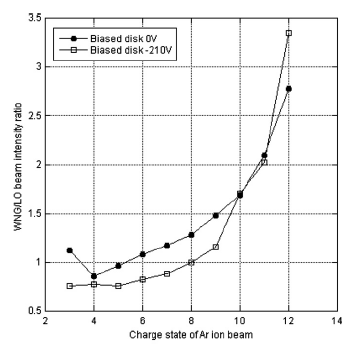}
\end{center}
\caption{Intensity ratios of argon ion beams generated with signals from the
WNG and LO versus the charge state\label{broad}}
\end{figure}

Some experiments have been performed with the JYFL 6.4 GHz
conventional minimum-$B$ geometry ECR ion source
at the University of Jyv\"askyl\"a, using Ar as the feed gas. The
performance of the source was compared when operated
with either narrow or with broadband microwave radiation,
under the same neutral gas pressure and input power
conditions. For the narrow-bandwidth experiments, the carrier signal
from the local oscillator (LO, bandwidth 1.5 MHz) was fed directly into
the TWTA. For the broadband experiments, the signal
from the LO was fed into a White-Noise Generator
(WNG), producing an output signal with a bandwidth of
200 MHz, FWHM, equally distributed about the central
frequency of the LO (6.4 GHz). The measurements were
carried out at the same absorbed microwave power level
(200 W), determined by subtracting the reflected power
from the forward power of the TWTA. The input power
was limited due to impedance mismatch of certain frequencies
in the amplified broadband signal, resulting in 20$\%$
reflected power of the injected power. This problem cannot be
avoided and it constitutes the main drawback of such a technique, since
frequencies are differently coupled into the ion source depending on the
intrinsic electromagnetic structure of the source itself. In the case of narrow-
bandwidth operation, the reflected power was typically 6$\%$. For the WNG
mode of operation at optimum pressure, the ion plasma
density, as evaluated from the measured drain current
and the charge-state distribution, using Bohm's criterion, was
within 15$\%$ of the neutral density. Although the beam intensities
were higher by factors >2 with the WNG at low pressures,
there was no difference between the broadband and
narrow-bandwidth modes of operation in the high neutral
pressure regime due to increased rates of charge exchange \cite{tarvainen,alton2}.
Enlarged ECR zones
improve the production of high-charge-state ions due to enhanced
bombardment with higher populations of energetic electrons
and, especially, due to the enhanced ionization rates of
neutrals in the ECR zone, thereby lowering the probability
of charge-exchange collisions. The tremendous potential
of the broadband technique for enhancing high-charge-state
beam intensities over conventional means at equivalent
input power levels is illustrated in Fig. \ref{broad}, where the
ratio of WN/LO-generated high-charge-state beams is seen
to increase continually with the charge state.
Clearly, the potential of the
broadband method can only be fully evaluated by comparing
the high charge states produced with the two modes of
operation at the same power density, if this is achievable in practice
in terms of the operational limitations imposed by outgassing
effects and the required RF power.

\begin{figure}[htpb]
\begin{center}
\includegraphics[width=0.5\textwidth]{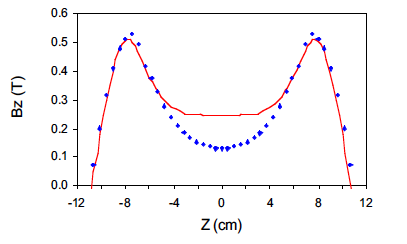}
\end{center}
\caption{Axial magnetic field profiles of the `volume'
(solid line) and conventional minimum-$B$ configurations
in `surface' (dotted line) ECR ion sources \cite{al4}. \label{Bflat}}
\end{figure}

The performance of ECR ion sources can also be significantly
improved by tailoring the central region of the magnetic
field so that it is resonant with single-frequency
microwave radiation (spatial domain) \cite{al1,al2,al3}.
The spatial-domain technique employs a magnetic
field configuration with an extended central flat region,
tuned to be in resonance with single-frequency microwave
radiation. Because of the large resonant plasma volume,
significantly more RF power can be coupled into the
plasma, resulting in heating of electrons over a much
larger volume than is possible in conventional ECR ion
sources. The ability to ionize a larger fraction of the particles in
the plasma volume effectively reduces the
probability of resonant and non-resonant charge exchange,
thereby increasing the residence time of an ion in a given
charge state and for subsequent and further ionization. All
other parameters being equal, the `volume' ECR source
should result in higher-charge-state distributions, higher
beam intensities, and improved operational stability.
The axial magnetic field is shown in Fig. \ref{Bflat}:
its profile is flat (constant
mod-$B$) in the centre, and extends over the length of the
central field region, along the axis of symmetry, and
radially outwards to form a uniformly distributed ECR
plasma `volume'. This magnetic field design strongly
contrasts with those used in conventional ECR ion
sources, where the central field regions are approximately
parabolic and the consequent ECR zones are `surfaces'.

According to computational studies, the new
configuration will result in dramatic increases in the
absorption of RF power, thus enabling the heating of
electrons over a much larger volume, thereby increasing
the electron temperature and the `hot' electron population in
the plasma.

\section{Frequency tuning}
Many experiments over recent years have shown that significant improvements
of ECRIS performance (both in terms of total extracted current
and the production of highly charged ions) can be obtained by slightly varying the
microwave frequency in the case of Single-Frequency Heating (SFH),
this being defined as a `frequency tuning' effect. It was already known that a large
increase in the frequency increases ($\sim$GHz) the electron density and improves
the ECRIS performance because of the increase of
the cut-off density, but since 2001 several experiments have demonstrated
that even slight variations of the pumping wave frequency ($\sim$MHz) may
lead to strong variations in the extracted current.
The first evidence was provided by the different performances observed for
the SERSE and CAESAR ion sources when fed by a klystron-based or a
 TWT-based generator \cite{lcel, gamm,von} at either 14 or
18 GHz.
Other interesting results came from experiments performed at GSI
and at JYFL.
In frequency tuning, the microwave frequency is varied
in order to select the most efficient heating mode inside the
plasma chamber of the ECRIS. In the experiments, the input frequency
for the klystron was swept from 14.05 to 14.13 GHz
in 100 s, using a Rohde $\&$ Schwartz signal generator. This bandwidth
was found adequate to maintain a constant output
power over the whole frequency sweep, using the automatic
level control feature of the klystron.

\subsection{Effect on beam structure and emittance}
\begin{figure}[htpb]
\begin{center}
\includegraphics[width=0.5\textwidth]{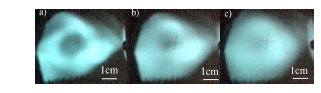}
\end{center}
\caption{(Colour online) The structure of the Ar$^{9+}$ ion beam with different
plasma heating frequencies is shown: (a) 14.050 GHz, (b) 14.090 GHz, and (c) 14.108 GHz.\label{emitt}}
\end{figure}

The beam emittance and the beam structure were studied
as a function of the frequency with the aid of an Allison-type
emittance scanner and a KBr beam viewer. Figure \ref{emitt} shows
three beam viewer pictures: the beam structure varies strongly with the microwave
frequency, as an indication of the change in the plasma--wave coupling
during the scan. However, no unequivocal
explanation concerning the origin of the beam structure variations
can be given: they could originate from the changes in
the plasma, in the electron ion dynamics, due to electromagnetic
field variations and/or in the beam line due to changes in
ion-beam intensity and space charge.
The explanation of the data presented above is based on the assumption that
the frequency tuning changes the electromagnetic field distribution inside the
resonator in terms of its distribution over the resonance surface; that is, where
the wave--electron energy transfer takes place. This assumption requires
that even in case of plasma filling the cavity, the resonant modes persist; that is,
the formation of standing waves is still possible.

\subsection{The microwave frequency dependence of the properties of the ion beam extracted from a
CAPRICE-type ECRIS}\label{Mic}
Measurements have
been performed with the CAPRICE-type ECRIS installed
at the ECR Injector Setup (EIS) of GSI. The experimental
set-up uses a microwave sweep generator
that feeds a TWTA, covering a wide frequency range from 8 to 18 GHz.
This arrangement provides a precise determination of the
frequencies and of the reflection coefficient along with
the beam properties. A sequence of viewing targets positioned
inside the beam line monitors the evolution of the beam shape.
In the present experiment, the frequency  tuning effect has been analysed in
the 12.5--16.5 GHz frequency range. The availability of a TWTA driven
by a signal generator made it possible to change the
source operating frequency in steps of a few hundred
kilohertz. This experiment allows us to analyse the beam properties
when the ECRIS operative frequency sweeps over a
wide range of 4 GHz, and hence for increasing ECR surfaces.
The influence on lower- and higher-charge states has
been analysed for different source conditions concerning
the magnetic field configuration, the gas pressure, and the
power setting.

\subsection{Description of experimental set-up}
The CAPRICE-type \cite{hitz} ECR ion source used for this
experiment is equipped with a 1.2 T maximum radial
magnetic field. The plasma chamber was 179
mm long and 64 mm in diameter. The RF power was
provided by a TWTA working in the 8--18 GHz frequency
range and able to provide an output power higher than
650 W in the frequency range of 12--18 GHz. The input of
the amplifier was driven by a signal generator able to
sweep from 1 to 20 GHz. According to the maximum
manageable power reflected to the amplifier, it has been
restricted to working at a power of 100 W and in the frequency
range of 12.5--16.5 GHz. The use of a waveguide
microwave isolator covering this frequency range and
handling up to 650 W could allow us to work with higher
powers. 

The frequency steps were set to 200 kHz, with a
dwell time of 20 ms for each step. The duration of
one measurement was then around 400 s. Two directional
couplers with high directivity were inserted in the
waveguide line in order to measure the forward power
and the reflected power by means of two microwave power
probes. The experiment was carried out with argon
and helium as support gases, at gas pressures of
(3.9--5.0)$\times10^{-6}$ mbar. The ion currents of the
Ar$^{7+}$, Ar$^{8+}$, and Ar$^{9+}$ charge states were measured
using a Faraday cup; the drain current of the high-voltage power supply of
the extraction was also recorded. The extraction voltage
was set to 15 kV; a voltage of $-2$ kV was applied to the
screening electrode. Viewing targets could be remotely
inserted at three positions along the beam line in order to
monitor the evolution of the beam shape right after the extraction,
the focused beam, and the analysed beam \cite{mader}. KBr was
used as the target coating material for this experiment.
\subsection{Results and discussion}

\begin{figure}[htpb]
\begin{center}
\includegraphics[width=1\textwidth]{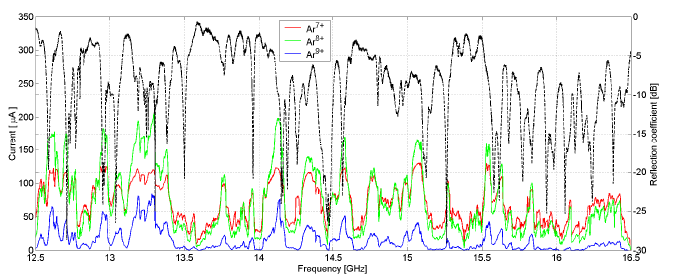}
\end{center}
\caption{(Colour online) The reflection coefficient and current evolution versus microwave frequency (the coloured solid lines refer to the left-hand
scale, while the dashed black line is the reflection coefficient, referring to the right-hand scale).\label{coeff}}
\end{figure}

\begin{figure}[htpb]
\begin{center}
\includegraphics[width=0.5\textwidth]{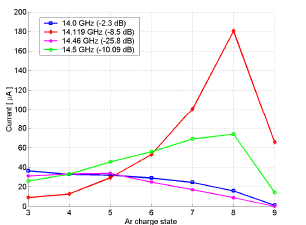}
\end{center}
\caption{The Ar charge-state distributions for four significant
operating frequencies\label{csd}}
\end{figure}

\begin{figure}[htpb]
\begin{center}
\includegraphics[width=0.5\textwidth]{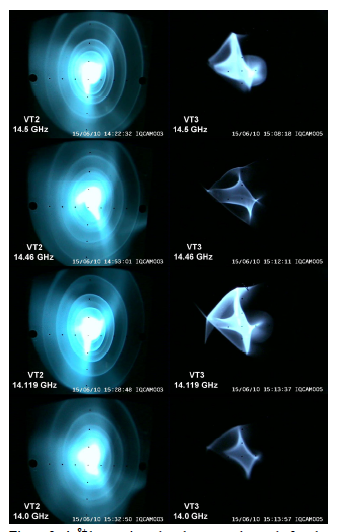}
\end{center}
\caption{The Ar$^{8+}$ beam viewed at the targets located after the
focusing solenoid (left-hand column) and after the dipole (right-hand
column).\label{em}}
\end{figure}

Different measurements were carried out by
sweeping the frequency and setting different ion source
parameters; that is, the injection and extraction magnetic field
values, the gas pressure, and the microwave power.
The source parameters were set to operate using a charge-state
distribution with a maximum on the Ar$^{8+}$ current
(by feeding the plasma with 100 W microwave power at 14.5 GHz).
An Ar$^{8+}$ current of 85 $\mu$A and a drain current of 2.36 mA
were obtained. From these source conditions, the frequency
sweep was started by ramping the signal generator
from 12.5 GHz up to 16.5 GHz, while the reflection coefficient,
the Ar$^{7+}$, Ar$^{8+}$, and Ar$^{9+}$ currents, and the
drain current were recorded simultaneously.
The evolution of the reflection coefficient with the frequency
is shown in Fig. \ref{coeff}.  As expected by
comparison with previous experiments, the matching impedance
between the cavity filled by the plasma and the electromagnetic wave is
strongly dependent on the frequency \cite{alton}. It is remarkable
that the plasma properties also change considerably
with varying frequency. The strong correlation
between the peaks of the reflection coefficient and the
current amplitude are clearly visible around the frequencies
at which the reflection coefficient is higher than $-9.54$
dB (matching condition). The relationship between the
resonance frequencies and the heating efficiency has been
analysed theoretically and particle-in-cell codes have been
used to correlate the electromagnetic field patterns and
the electron cyclotron resonance surface \cite{david}. However, it
is not possible with our analysis to determine the electromagnetic
field patterns (modes) related to these peaks.
The comparison of the current evolution in the frequency
range indicated above is presented in Fig. \ref{coeff}.
It is clear
how the current amplitude is affected by the choice of the
operative frequency. Looking at the Ar$^{8+}$ current, it ranges
from a few microamperes up to 200 $\mu$A. The experimental results
were clearly reproducible in several runs, thus confirming
the reliability of the measurements. The evolution of the
Ar$^{7+}$ and Ar$^{8+}$ currents is similar, but their amplitudes differ.
In fact, at the frequencies at which both currents present
a peak, the Ar$^{8+}$ current is quite a bit higher than the Ar$^{7+}$ current (e.g.
at 14.119 GHz, the difference between the two currents
is 80 $\mu$A). The opposite behaviour is visible for the
minima of the current amplitudes, where the Ar$^{7+}$ current is higher
than the Ar$^{8+}$ current. In the range 15.64--16.5 GHz, the currents of
the higher-charge states -- that is, Ar$^{9+}$ -- tend to lower values
even if the reflected power is less than 10$\%$. This seems
to be due to the confining magnetic field, which restricts the
source operation to lower frequencies. It has been also
observed that the evolution of the drain current follows the
trend of the three charge states presented in Fig. \ref{coeff}.
In order to have a complete understanding of the
sweep effects on the ionization process, the charge-state
distribution has been analysed for different frequencies. It
was decided to restrict the analysis to the frequency
range of 14--15 GHz. Several frequencies have been considered
at which peaks and minimum amplitudes occur. In
Fig. \ref{csd}, the charge-state distributions are presented for four
different frequencies.
The 14.5 GHz value is the normal
operational frequency of the CAPRICE ion source; 14.46
GHz and 14.119 GHz are the frequencies at which
the minimum and the maximum Ar$^{8+}$ currents, respectively, were
measured in the 14--15 GHz frequency range. The charge-state
distribution related to 14.0 GHz operation is also
reported in order to emphasize the importance of the electromagnetic
field pattern and the choice of the frequency.  In
fact, the charge-state distributions at 14.0 GHz and at
14.46 GHz are quite similar, even though the amount of power
feeding the plasma was more than doubled. In the first
case, the reflection coefficient (indicated in the legend of
Fig. \ref{broad}) was $-2.3$ dB and more than half of the power
was reflected; and in the second case it was $-25.8$ dB, the
impedance matching condition was fulfilled, and hardly
any power was reflected. It is also interesting that for the
frequencies at which the higher-charge states are favoured,
the current of the lower charge state is decreasing and vice
versa. The analysed charge-state distributions confirm that
frequency tuning affects the higher-charge
states to a greater extent (at 14.119 GHz, the current enhancement with respect to
the 14.5 GHz operational frequency is 244$\%$ for the Ar$^{8+}$ current and
456$\%$ for the Ar$^{9+}$ current).
The enhancement of the current is not the only effect of
frequency sweeping; in fact, the quality, the shape,
and the emittance of the ion beam also vary. The use of
beam viewing targets has proven to be a favourable technique
for monitoring the beam shape and a promising beam
diagnostic tool. The images recorded beyond the extracted
beam-focusing solenoid (position VT2) and in the diagnostic
box after the mass/charge selection dipole (position VT3)
are shown in Fig. \ref{em}.

Here, the same frequencies
are chosen as in Fig. \ref{csd}. The Ar$^{8+}$ beam is shown in
the right column and the focused beam images also refer
to the Ar$^{8+}$ magnetic field setting. At 14.0 GHz and 14.46
GHz, which are the frequencies at which the Ar$^{8+}$ current
presents a minimum, the focused beam shape seems to remain
unchanged in the orientation of the arms (at 14.0 GHz it is a
little bit brighter, also according to the higher intensity
measured at the Faraday cup). At 14.119 GHz, the focused
beam is a little bit bigger and brighter than the one at 14.5
GHz, and in both cases the orientation of the arms is
turned by more then 40$^\circ$ clockwise with respect to
the 14--14.6 GHz beam shapes.
 \subsection {Frequency tuning combined with the two-frequency heating effect}
The relationship between the two frequencies and the respective power
was not univocally determined. In fact, any source features a different set of
parameters and the optimization is done empirically,
just by looking at the maximization of the beam current. Several qualitative
explanations have been offered for this phenomenon, all related to the
increase of the average electron temperature, T$_e$, and to the ionization rate, by
assuming that the crossing of two resonance surfaces helps the electrons to
gain more energy. This simple picture does not explain the reason for the relevant
changes in the charge-state distribution for different pairs
of frequencies (even for the case of minor changes -- such as a few megahertz
over 14 or 18 GHz), which can be explained nowadays in terms of the frequency
tuning effect. It is important to underline that, even in the case of TFH
applied to many existing sources, a TWT is often used, the other option being a
klystron-based generator. The choice of a TWT allows the experimentalists
to vary the second frequency slightly. It has been observed -- for example, for
the production of O$^{7+}$ -- that 60 W emitted by the TWT at the optimum
frequency gives the same effect as 300 W emitted from the
fixed-frequency klystron. The maximum current is obtained by means of a
klystron at 427 W and a TWT at 62 W ($I$ = 66 $\mu$A), operating simultaneously;
in order to obtain the same current, the experimenters needed $P_{\mathrm{RF}}$ = 800 W
from the klystron in SFH. Furthermore, in the case of TFH, the current increases almost
20$\%$ (from 57 to 66 $\mu$eA) when the TWT-emitted frequency shifts from
11.06 to 10.85 GHz,  the klystron and TWT emitted power both being held constant.
Then, the TFH is an effective method to increase the extracted current from an
ECRIS, but it can only be fully exploited by means of frequency tuning.
Several measurements have been carried out with the SERSE ion source.
The TFH has been used for operation at either 14 or 18 GHz, with a clear
advantage with respect to SFH.

\begin{figure}[htpb]
\begin{center}
\includegraphics[width=0.5\textwidth]{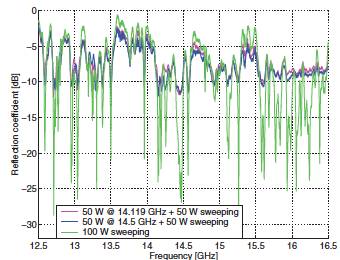}
\end{center}
\caption{(Colour online) The reflection coefficient as a function of frequency: a
comparison between the frequency tuning effect and double-frequency
heating.\label{dfh}}
\end{figure}

\begin{figure}[htpb]
\begin{center}
\includegraphics[width=0.5\textwidth]{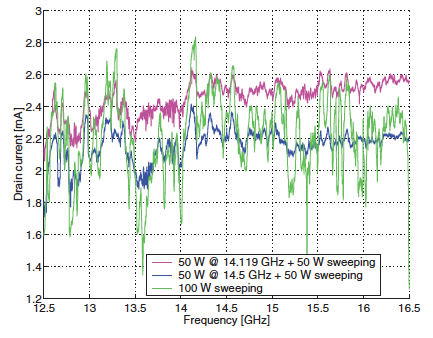}
\end{center}
\caption{(Colour online) The drain current of the high-voltage extractor power
supply as a function of frequency: a comparison between the frequency tuning
effect and double-frequency heating.\label{drain}}
\end{figure}

\begin{figure}[htpb]
\begin{center}
\includegraphics[width=0.5\textwidth]{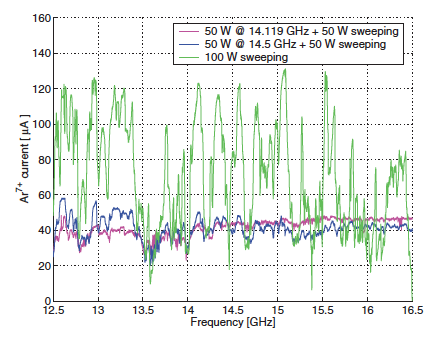}
\end{center}
\caption{(Colour online) The Ar$^{7+}$ current as a function of frequency: a comparison
between the frequency tuning effect and double-frequency heating.\label{artf}}
\end{figure}

\begin{figure}[htpb]
\begin{center}
\includegraphics[width=0.5\textwidth]{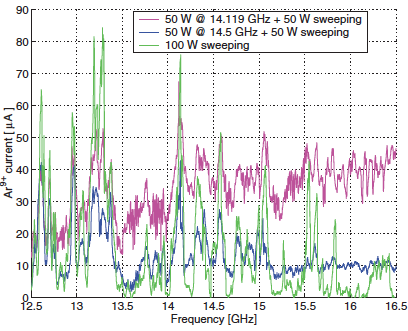}
\end{center}
\caption{(Colour online) The Ar$^{9+}$ current as a function of frequency: a comparison
between the frequency tuning effect and double-frequency heating.\label{ar9f}}
\end{figure}

\begin{figure}[htpb]
\begin{center}
\includegraphics[width=0.5\textwidth]{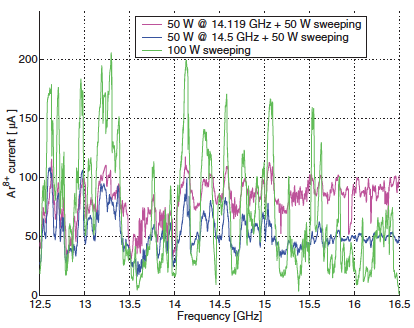}
\end{center}
\caption{(Colour online) The Ar$^{8+}$ current as a function of frequency: a comparison
between the frequency tuning effect and double-frequency heating.\label{ar8f}}
\end{figure}

In section \ref{Mic}, the evolution of the ion current intensities
of Ar$^{7+}$, Ar$^{8+}$, and Ar$^{9+}$ over the 12.5--16.5 GHz frequency
range has been described. The argon charge-state distributions
were analysed for different frequencies, and for some
frequencies an enhanced intensity of the higher-charge states
was observed. We decided to perform
an experiment to investigate the double-frequency heating effect
using a single fixed frequency (14.5 GHz in one case
and 14.119 GHz in another) and a second frequency swept
over the 12.5--16.5 GHz range. First, the power feeding the
plasma was kept at 100 W, in order to compare the results
of double-frequency heating with single-frequency sweeping.
Then the power distribution between the fixed frequency
and the sweeping frequency was unbalanced by doubling the
power of the fixed frequency in one case and the power of
the sweeping frequency in the other. The ion source
parameters -- including the magnetic field, the gas pressure, the extraction
voltage, and so on -- were set to and maintained at the same values
as used for single-frequency sweep analysis, with the charge-state
distribution optimized for the Ar$^{8+}$ intensity. The Ar$^{7+}$,
Ar$^{8+}$, and Ar$^{9+}$ currents were recorded together with the extractor
drain current. The forward power and the reflected
power were measured simultaneously. However, in the case of
double-frequency heating, the power probes cannot distinguish
the measured values of the forward and reflected power related to
each of the two waves. The results of these measurements are shown
in Fig. \ref{dfh}.
 A comparison between the two techniques
concerning the behaviour of the ion source is possible.
As expected, in the case of the two frequencies, the minima of
the reflection coefficient are higher because of the higher level
of reflected power due to the superposition of the two waves.
It is interesting to note that these minima occur at almost the
same frequencies for both techniques.
Since the positions of the reflection coefficient minima do
not vary, it can be expected that the peaks of the ion current
are located at the same frequencies for the two heating methods.
This comparison is shown in Fig. \ref{drain} for the drain current
of the high-voltage extractor power supply. In this case, 50 W
is provided at one fixed frequency and the effect of sweeping
a second frequency with a power of 50 W produces a different
current behaviour from that with a single sweeping frequency (at
the same power). This can be seen in Fig. \ref{drain} and Figs. \ref{artf}--\ref{ar9f}.
In fact, during the single-frequency sweep, the drain current
varies from around 1.4 mA to 2.8 mA, while with double-frequency
heating the range of variation is 0.6 mA. The
drain current appears more stable with frequency when the
double frequency is applied instead of single-frequency tuning.
This behaviour is also observed in the Ar$^{7+}$ current variation
shown in Fig. \ref{artf}. A possible explanation might be found
in the microwave power provided by the two methods. In the
case of single-frequency tuning, 100 W is provided to the
source and during the frequency sweep, when the impedance
is not matched, most of the forward power is not coupled to
the source but is reflected (see Fig. \ref{dfh}). But in the case of
double-frequency heating, just 50 W of the total forward power
of 100 W is associated with the sweeping frequency, and at
the frequencies at which the power is not coupled to the plasma,
one half of the total power is provided at a fixed frequency,
which is well coupled to the source (i.e. at 14.5 GHz or at
an optimized frequency such as 14.119 GHz). The choice of the
fixed frequency allows enhancement of the ion currents of the
higher-charge states. With respect to single-frequency operation, with two frequencies the current
peaks are broadened and reduced in amplitude, the frequencies
at which they occur remain the same, and the average
current level remains high (see Fig. \ref{artf} for Ar$^{7+}$), while the
higher-charge states tend to higher ion currents when the frequency
is increased (see Figs. \ref{ar9f} and \ref{ar8f}). The experimental
results indicate that there can be strong
differences when the operating frequency is optimized by frequency
tuning. In fact, when operating in double-frequency
mode and optimizing one of the two frequencies (14.119 GHz
for the present case, as described above), the current increases
linearly with frequency (with respect to the 14.5 GHz fixed
frequency).

\begin{figure}[htpb]
\begin{center}
\includegraphics[width=0.5\textwidth]{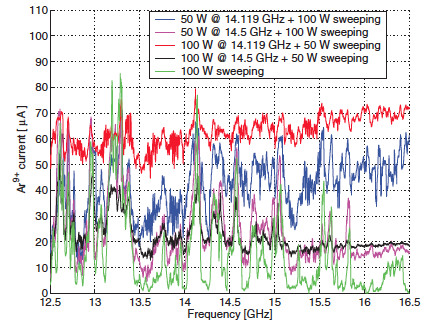}
\end{center}
\caption{(Colour online) The Ar$^{9+}$ current as a function of frequency: differing
power distributions between the two waves.\label{curr}}
\end{figure}

In comparing the two techniques while keeping the ion source
settings and the microwave power constant,
note that the behaviour of the ion currents and the reflection coefficient
remain almost the same with frequency. This result is
useful to optimize the performance of an ECRIS further when
double-frequency heating is used. In fact, single-frequency tuning
can be used to find the optimum frequencies to be used for
operating in the multiple heating mode. Note also that when
operating with two generators set at the same frequency, the
current level measured is different from the single-generator
case. This is because when the two frequencies become equal,
the current of each charge state and the power provided to the
ion source are not stable. A possible explanation for this phenomenon
might be overlapping of the two signals and the
phase difference between the two waves. We plan to explore
the use of two phase-locked generators in forthcoming experiments.
The behaviour of argon ion beams has been explored for different
power levels of the fixed frequency and the sweep frequency.
In both cases, two fixed frequencies were chosen and
in this case the possibility of setting the fixed frequency to an optimized
value, obtained by fine-tuning the frequency, allowed
us to increase the beam intensity (Fig. \ref{curr}).

Figure 13 shows the evolution of the Ar$^{9+}$ current
when 150 W of total power is distributed between the two microwave
feeds in different proportions, one of fixed frequency and the
second swept from 12.5 GHz to 16.5 GHz. Again, the
single-frequency sweep carried out at 100 W is included in the figure
for comparison. For the optimum frequency mentioned previously
(14.119 GHz), the current intensity is still comparable between
the two methods even though the total power is 50$\%$ greater.

\begin{figure}[htpb]
\begin{center}
\includegraphics[width=0.5\textwidth]{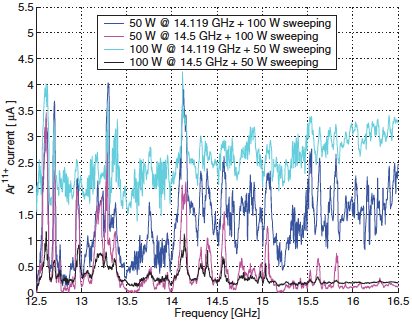}
\end{center}
\caption{(Colour online) The Ar$^{11+}$ current as a function of frequency, showing different
power distributions between the two waves.\label{ar11}}
\end{figure}

When the power associated with the fixed frequency is higher,
all the analysed charge states are less sensitive to variation
of the second frequency. This follows from the reduced dynamic
range of the current, indicating a more stable plasma;
in particular, when the fixed frequency is chosen to be at the
optimum value (14.119 GHz). Furthermore, when one of the
two frequencies is optimized, the average extracted current
remains high with respect to an unoptimized frequency (such
as 14.5 GHz), increasing with the second frequency (quite evident
above 15 GHz). All these observations are more evident
for higher-charge states.
With the ion source running at higher microwave power,
it was possible to perform an experiment with an Ar$^{11+}$ ion
beam. The results are shown in Fig. \ref{ar11}.

 The power distribution
between the two waves is also important. Obviously,
it is more effective to increase the power at the fixed frequency.
Figure \ref{ar11} shows once more that the choice of the optimum
frequency, as obtained in the previous experiment on
frequency tuning, is very important for double-frequency heating
at higher power. This result is interesting with respect to
the operation of new-generation ECRIS machines working at 28 GHz,
using gyrotron microwave generators in combination with a
second microwave generator of lower frequency, working in
double-frequency mode.

\section{Alternative heating method: Bernstein waves}
 ECR devices are density limited, because the electromagnetic waves cannot propagate beyond a certain density, called the cut-off density.
An alternative to the classical ECR interaction is electrostatic wave heating, driven by Bernstein waves. Bernstein waves (BW) are very interesting for
nuclear fusion devices \cite{bern1}, because the plasma heating occurs in the absence of any density
cut-off. Electromagnetic Ordinary (O) or Extraordinary (X) modes can be externally
launched into the plasma.
O-mode waves may be coupled to an X mode in the O-mode cut-off layer, and then the
X mode can be coupled to a BW mode at the upper hybrid resonance (UHR). This process is called
O-X-B mode conversion and was described in 1973 by Preinhaelter and Kopecky \cite{bern1}.

\begin{figure}[htpb]
\begin{center}
\includegraphics[width=0.5\textwidth]{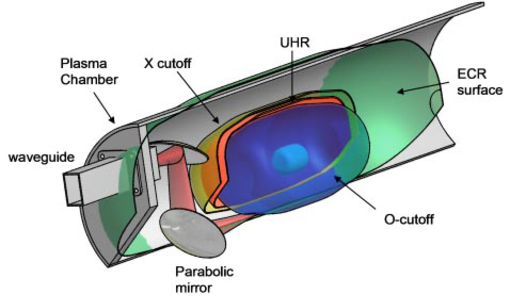}
\end{center}
\caption{A possible design for a microwave injection system producing O-X-B conversion: two parabolic mirrors are used to bring a focused microwave beam into the optimal region of the O cut-off layer.\label{b}}
\end{figure}

It was shown that O-X-B mode conversion may be optimized by changing the O-mode
insertion angle with respect to the external magnetic field direction: in this case, the O
mode is completely converted into a slow X mode, which under certain conditions is in turn converted to
BWs in the Upper Hybrid Resonance (UHR) layer \cite{pres}. The generated Bernstein waves travel
inside the plasma until they are absorbed at the ECR or at the higher-order cyclotron
harmonics.
A possible design for a microwave injection system producing the O-X-B conversion is shown in Fig. \ref{b}: first, an O wave is launched from the outside, with an oblique angle of incidence obtained with a proper orientation of the parabolic mirrors. For an optimal launch angle, it is possible to obtain a correlated optimal parallel refractive index and then a coincidence of the O and X modes at the critical plasma density (cut-off). This
means that both modes have the same phase and group velocities and the power is transferred
without reflections.
Once the X waves are generated, they propagate towards the UHR. Here, the X mode
coincides with the electron Bernstein mode. In the linear description, the X waves are
completely converted into EBWs. This process is called X--B conversion. It should be
noted that the O-X-B process can only take place if the plasma density is above the O-wave
cut-off density.

O-X-B plasma heating and current driving with BW in an overdense plasma were
demonstrated in the Stellarator WEGA, operating at the Max Planck Institute for Plasma
Physics in Greifswald, Germany \cite{bern2}.
The heating of a plasma by means of EBW at particular frequencies enabled us to reach densities much larger than the cut-off ones. Evidence of EBW generation and absorption together with X-ray emissions due to high-energy electrons is shown in ray tracing simulations and CCD photographs in Ref. \cite{cas}.

A plasma reactor operating at the Laboratori Nazionali del Sud of INFN, Catania, has been used as a test-bench for the investigation of innovative mechanisms of plasma ignition based on electrostatic waves (ES-W), obtained via the inner plasma EM-to-ES wave conversion. Evidence of Bernstein wave (BW) generation is shown in Ref. \cite{bern3}.

In particular, during the experiment, a microwave discharge
ion source has been used: a plasma reactor consisting of a stainless-steel cylinder that is 24 cm long and 14 cm in diameter. A NdFeB permanent magnet system generates
an off-resonance magnetic field along the plasma chamber
axis (with a maximum of 0.1 T on axis). 

\begin{figure}[htpb]
\begin{center}
\includegraphics[width=0.5\textwidth]{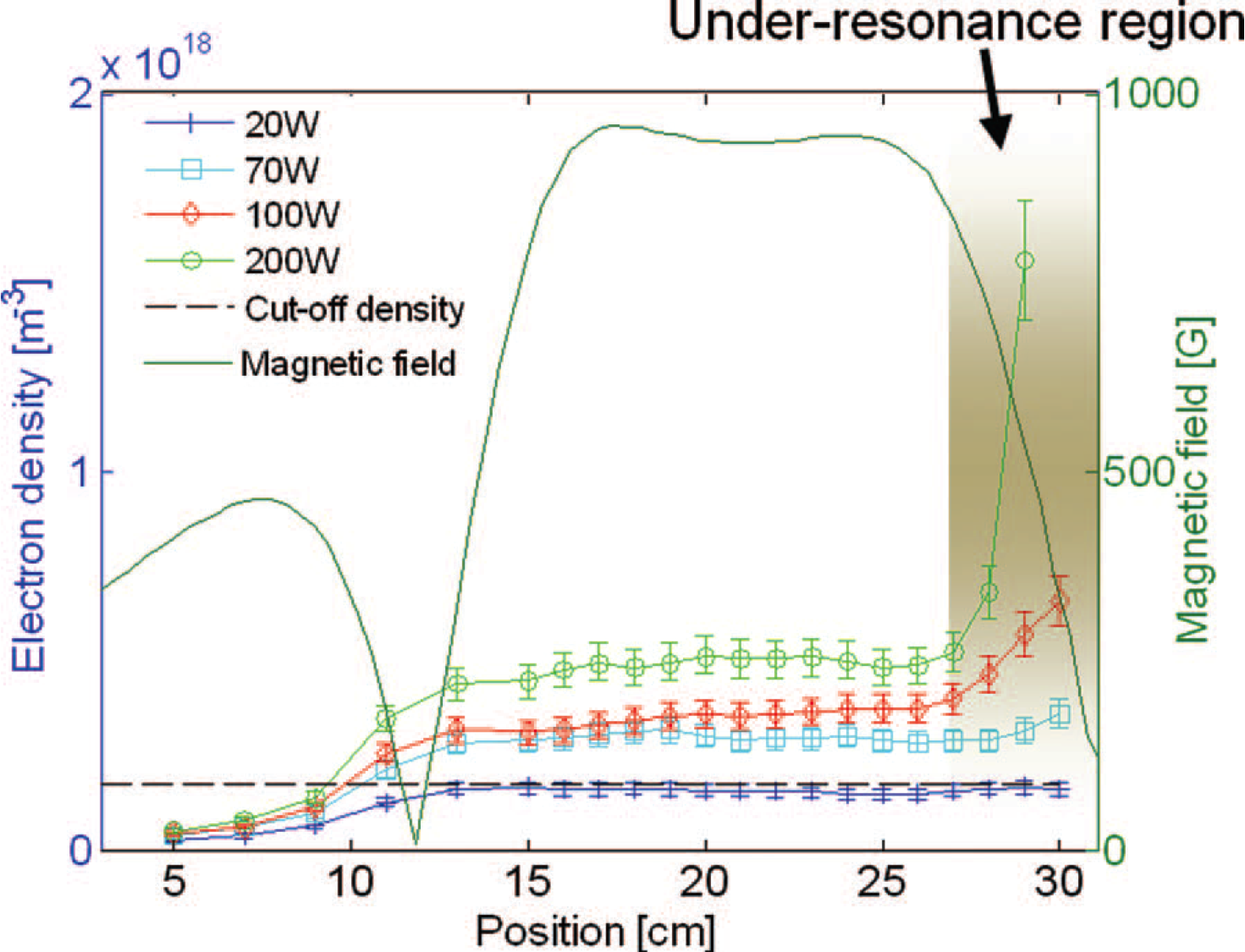}
\end{center}
\caption{(Colour online) Electron density and magnetic field profile at
$1.5\times10^{-4}$ mbar, and frequency 2.45 GHz, at different microwave powers.
Microwave injection occurs at the right-hand side of the figure.\label{den}}
\end{figure}

Microwaves have been generated by using a TWT,
which is able to generate microwaves from 3.2 to 4.9 GHz. The typical
working frequency when using the TWT was 3.7478 GHz.
The temperature and plasma density measurements have
been carried out by using a movable Langmuir Probe (LP).
The LP can host a small wire used as a local electromagnetic
antenna, which can be connected to a spectrum analyser
for the plasma spectral emission analysis. An Si-Pin and a
HPGe X-ray detector have been used for the measurement of
X-ray spectra in different plasma conditions. Both detectors
are able to detect X rays with energies greater than about 1
keV. A CCD camera has been used to visualize the plasma
structure within the chamber at different working frequencies, and
at different microwave powers and pressures.
A series of LP measurements has been carried out with the plasma reactor at the frequency
of 2.45 GHz, when both under-resonance and off-resonance
regions are present. In Fig. \ref{den}, it is evident that the electron
density is drastically enhanced in regions where the condition $B
< B_{\mathrm{ECR}}$ is satisfied.

\begin{figure}[htpb]
\begin{center}
\includegraphics[width=0.5\textwidth]{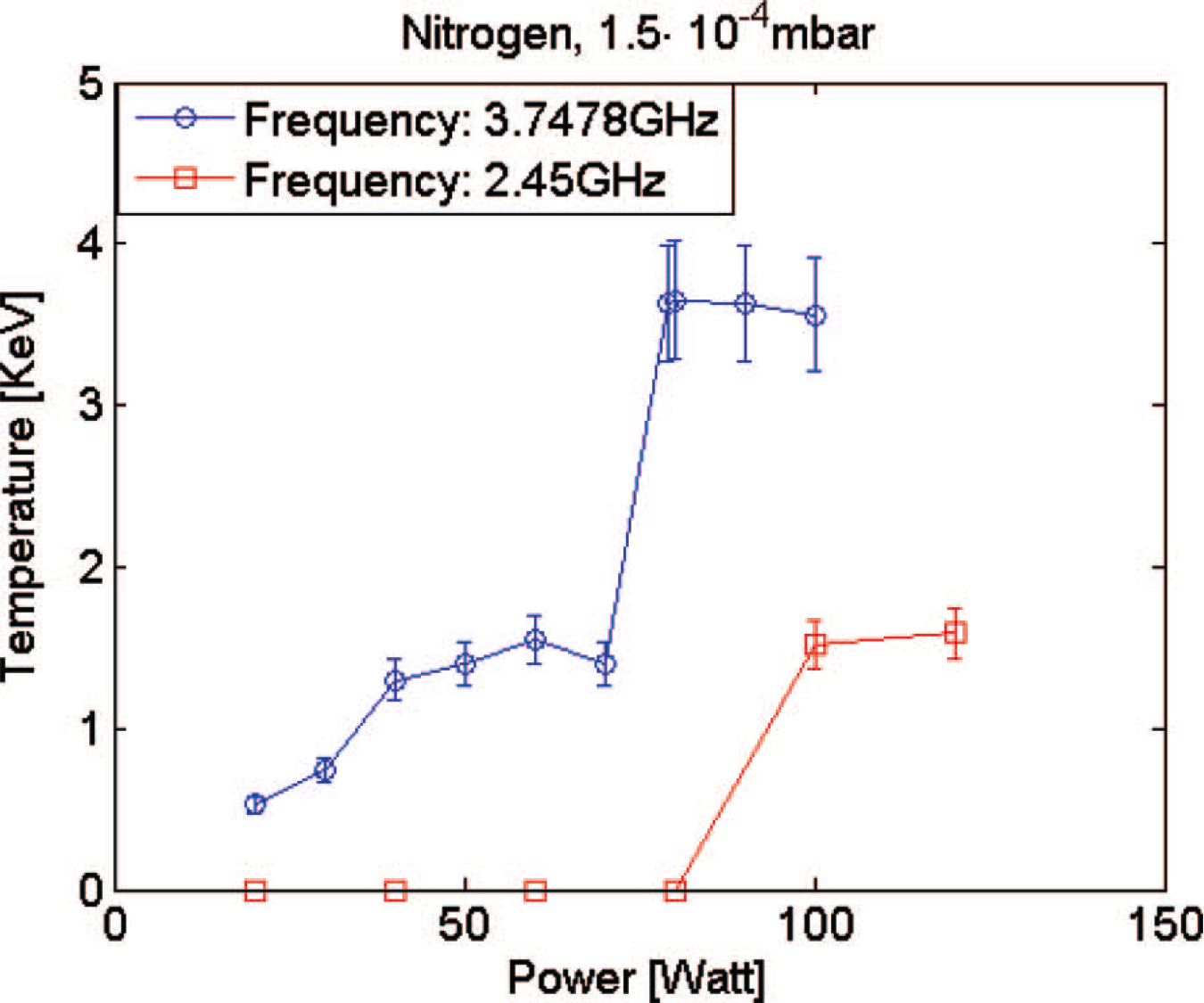}
\end{center}
\caption{(Colour online) Spectral temperatures. At a frequency of 2.45 GHz,
ECR heating and EBW heating coexist; at 3.7478 GHz, EBW heating is
dominant.\label{tem}}
\end{figure}

This effect is observed for all of the different
power values that we used; in particular, at 200 W, an electron
density of about $1.5\times10^{12}$ cm$^{−3}$
 has been measured, a value 20 times
greater than the cut-off density. Note that the electron
density is everywhere comparable or larger than the cut-off
density ($n_c = 7.5\times10^{10}$ cm$^{−3}$).
In the same magnetic configuration, X-ray measurements
have been carried out at 2.45 GHz and 3.7478 GHz. By increasing
the pumping frequency, it is possible totally to remove
the ECR, so that at 3.7478 GHz the entire plasma chamber
is at under-resonance and EBW heating becomes the unique
heating mechanism. Spectral temperatures measured in the
two cases are shown in Fig. \ref{tem}.

 At the frequency of 3.7478
GHz, the spectral temperature is significantly larger (up to 4
keV) than at 2.45 GHz. Around 80 W, it is possible to identify
a threshold for which the temperature slope increases
suddenly and becomes steeper for both configurations.
The end-point energy is about 10 times the value of the spectral
temperature. The spectral temperature decreases slightly
with the atomic mass of the gas used and when the pressure is
increased.  Further results and discussion can be found in Ref. \cite{cas}.
The results are interpreted through the Bernstein wave heating theory, and
are very promising for future high-intensity multicharged ion sources that
can, therefore, be based on the results described here, by employing a
simplified magnetic configuration with respect to typical minimum-$B$ ECR
ion sources.

\section{Conclusion}
Microwave coupling to ECR ion sources plays a fundamental role in enhancing their performance in terms of current extracted and average charge state produced.
Over the past decade, new heating methods have been studied to overcome the technological limitations due to the magnetic field and frequency scaling, to achieve the production of milliampere levels of HCI, as requested for the new accelerating facilities all over the world.

\end{document}